# COPPER PLANAR MICROCOILS APPLIED TO MAGNETIC ACTUATION

*Johan MOULIN, Marion WOYTASIK, Emile MARTINCIC, Elisabeth DUFOUR-GERGAM*

IEF, Université Paris XI, 91 405 Orsay Cedex, France

## ABSTRACT

Recent advances in microtechnology allow realization of planar microcoils. These components are integrated in MEMS as magnetic sensor or actuator. In the latter case, it is necessary to maximize the effective magnetic field which is proportional to the current passing through the copper track and depends on the distance to the generation microcoil.

The aim of this work was to determine the optimal microcoil design configuration for magnetic field generation. The results were applied to magnetic actuation, taking into account technological constraints. In particular, we have considered different realistic configurations that involve a magnetically actuated device coupled to a microcoil. Calculations by a semi-analytical method using Matlab software were validated by experimental measurements.

The copper planar microcoils are fabricated by U.V. micromoulding on different substrates: flexible polymer (Kapton®) and silicate on silicon. They are constituted by a spiral-like continuous track. Their total surface is about 1 mm$^2$.

## 1. INTRODUCTION

The development of mechanical microsystems that require a high amplitude motion is dependant on the advances in long range and high density energy transmission systems. Both new magnetic materials and technology for microcoil are studied for improving generation of high magnetic fields. In the later case, planar microcoils are extensively used for actuation of relays [1], membranes for pump or valve [2, 3] or flexible beams [4].

In addition to long range and high density energy, other advantages of magnetic actuation are the facility to process arrays of microcoils [5, 6] and to improve the magnetic field by mutuality of several coils. An important point is that magnetic actuation is suitable for biomedical applications, as low frequency magnetic field is not dangerous for biological tissues. On the other hands, the main disadvantage of the magnetic actuation is the electrical consumption, due to Joule losses.

The present study aims to define the more effective configuration of planar microcoils in terms of magnetic field only, then taking into account the electrical consumption. After the realization of several demonstrators, experimental measurements were performed. As thermal properties of the substrate and packaging lead the maximum current in the coil, different realistic configurations have been proposed. The maximum magnetic field has been evaluated for each of them.

## 2. MICROCOILS EFFICIENCY FOR MAGNETIC FIELD GENERATION

### 2.1 Fabrication process

The fabrication process of the planar microcoils is based on the U.V. micromoulding technology (fig 1). The two main steps are the optical lithography and the copper electroplating through photoresist moulds. This process is of great interest since it allows a fast, low cost realization of thick metallic tracks. Moreover, one mask level only is necessary. The microcoils can be processed on different substrates: flexible polymer (such as Kapton®) and silicate on silicon. It is to notice that the process is the same for both substrate types, except for the pretreatment of the Kapton® [7].

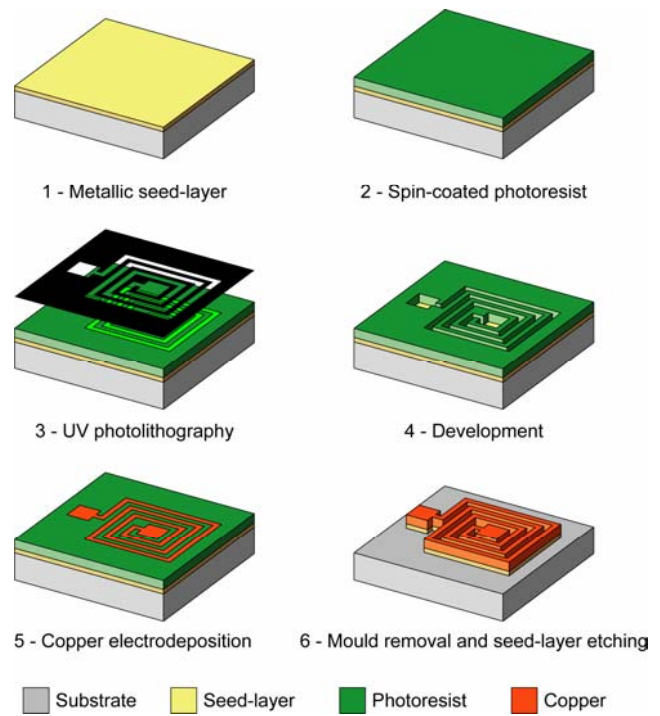

Figure 1: Process technology steps for microcoil fabrication

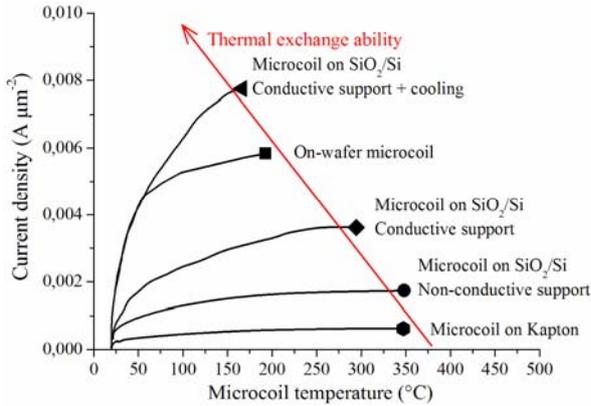

Figure 2: Influence of the packaging on the maximum current density in the track of a microcoil

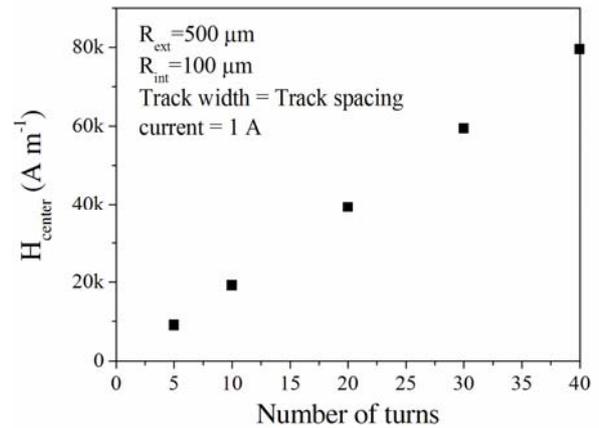

Figure 3: Magnetic field at the center of the microcoil

Different shapes were defined: round (R) and square (S). The track width and spacing are variable but wider than 5 µm. Compared to the mask dimensions the optical lithography can induce small variations. The number of turns is in the 1-40 range. Nevertheless, the external dimension of the coils is 1 mm. The track thickness is predetermined by the electrodeposition conditions (time and current density) but is limited by the photoresist one, i.e. 20 µm.

The microcoils were glued on different supports and their connecting pads were microbonded with a 25 µm diameter aluminum wire.

The track thickness and width were precisely defined using optical interferometry and SEM observations respectively.

### 2.2 Limit current

The current in the microcoils is limited by Joule effect. It has been established in previous work that the thermal exchange ability of the substrate drives the maximal current density in the coil [8]. For instance, current density is limited to 0.6 mA µm$^{-2}$ in microcoils processed on Kapton® and to 6 mA µm$^{-2}$ in on-wafer microcoils (see fig. 2).

### 2.3 Effect of the geometry

The magnetic field generated by a microcoil is proportional to the injected current and increases with its number of turns.

One considers a circular track indexed by n crossed by a uniform current I. If its radius varies from $R_{n\,min}$ to $R_{n\,max}$, it generates at its center a magnetic field:

$$H_n = \frac{I}{2(R_{n\,max} - R_{n\,min})} \ln\left(\frac{R_{n\,max}}{R_{n\,min}}\right) \quad (1)$$

The expression of the magnetic field generated by the N turns of the microcoil is:

$$H_{center} = \sum_{n=1}^{N} H_n \quad (2)$$

By noting w and s the track width and spacing respectively, the iteration related to the radii are:

$$R_{n\,max} = R_{n\,min} + w \quad (3)$$

and

$$R_{n\,min} = R_{min} + (n-1)(w+s) \quad (4)$$

By applying this relation to n = N, the relation between the internal and external radii of the microcoil $R_{min}$ and $R_{max}$ is:

$$R_{max} - R_{min} = (N-1)(w+s) + w \quad (5)$$

Thus, using formulae (2) and (5), the expression of the magnetic field generated by the microcoil is:

$$H_{center} = \frac{I}{2w} \sum_{n=1}^{N} \ln\left(1 + \frac{w}{R_{max} - (N-n)(w+s) - w}\right) \quad (6)$$

The figure 3 presents the computing of $H_{center}$ in the following configurations: round microcoils with $R_{min}$ = 0.1, $R_{max}$ = 0.5, w = s and the number of turns N varying from 5 to 40. For instance, w = s = 5 µm for the N = 40 turns coil. The computation shows that $H_{center}$ is proportional to the number of turns N.

However, if N increases, the cross-section of the track decreases so as the maximum current that can be injected into the track. The Maximum Effective Magnetic Field (M.E.M.F.) at the center of the microcoil is equal to the product of the magnetic field related to geometry for I = 1 A by the maximal current allowed in the track (see 2.2).

*Johan MOULIN, Marion WOYTASIK, Emile MARTINCIC, Elisabeth DUFOUR-GERGAM*
*Copper Planar Microcoils Applied to Magnetic Actuation*

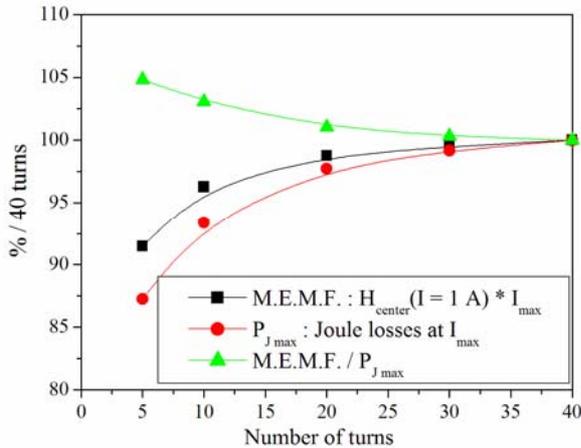

Figure 4: Maximum Effective Magnetic Field (M.E.M.F.) of the coil, corresponding Joule losses and their ratio as a function of the number of turns (Normalized to 40 turns)

The figure 4 shows that the M.E.M.F. of the microcoil increases lightly with the number of turns. Its value for a 5 turns coil is 92 % of that of a 40 turns coil.

Finally, in order to take into account the electrical consumption of the device, the Joule losses $P_{J\ max}$ corresponding to the maximum current in the coil have been computed, using the relation:

$$P_{J\ max} = RI_{max}^2 = \rho_e \frac{\ell}{S} j_{max}^2 S^2 = \rho_e \ell\, j_{max}^2 S \qquad (7)$$

with $\rho_e = 1{,}7\ 10^{-8}\ \Omega m$ the electrical resistivity of the electrodeposited copper, S the cross-section of the track and $\ell$ the mean length of the coil:

$$\ell = \pi\left[2NR_{max} - Nw - (w+s)(N-1)(N+2)\right] \qquad (8)$$

The ratio M.E.M.F./$P_{J\ max}$ represents the efficiency of the microcoil for magnetic field generation taking into account the electrical consumption. Its values have been calculated as a function of N and normalized to N= 40 turns in the figure 4. As a reference, the maximum current in a 40 turns coil which track thickness is 10 μm, processed on silicon and glued on a TO220 support is about 175 mA (see fig. 2). So the M.E.M.F., $P_{J\ max}$ and their ration are equal to 14000 A m$^{-1}$, 0.76 W and 18000 A m$^{-1}$ W$^{-1}$ respectively. The results show that if the 40 turns coil corresponds to the more efficient configuration in terms of magnetic field generation, it is although the more consuming in terms of energy (around 5% more than a 5 turns coil).

### 2.4 Spatial characteristics

The magnetic field generated on the axe of two round and square 40 turns planar microcoils (w = s = 5 μm) has been computed using the following formulae:

$$H_{round}(d) = \frac{I}{2}\sum_{n=1}^{N}\frac{R_n^2}{\left(d^2 + R_n^2\right)^{\frac{3}{2}}} \qquad (9)$$

and

$$H_{square} = \frac{2I}{\pi}\sum_{n=1}^{N}\frac{R_n}{\left(d^2 + R_n^2\right)}\sin\left(a\tan\left(\frac{R_n}{\sqrt{d^2 + R_n^2}}\right)\right) \qquad (10)$$

with d the distance to the coil center.

The calculation supposes that the width of the track is negligible, which is appropriate for a 40 turns microcoil. In addition, the current was chosen to be 300 mA.

As shown on the figure 5, the magnetic field decreases drastically if the distance to the coil is larger than several tens of microns.

In addition, a round coil is around 10% more efficient than a square one for magnetic field generation, until the distance to the coil remains below 100 μm.

### 2.5 Experimental characterization

The magnetic field generated by round and square 40 turn microcoils has been experimentally measured using two high sensitive sensors: a Honeywell HMC 1001 magneto-resistance and an Aichi Micro Intelligence Corporation magneto-impedance. The sensitivity and the spatial resolution of the first are lower than that of the latter. However, the distance to coil is closer for the MR sensor than for the MI one. That explains the different ranges of measured magnetic field.

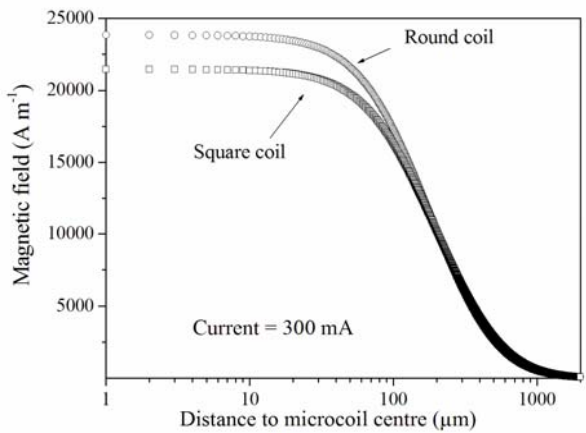

Figure 5: Magnetic field generated by two round and square 40 turn planar microcoils supplied by a 300 mA current.

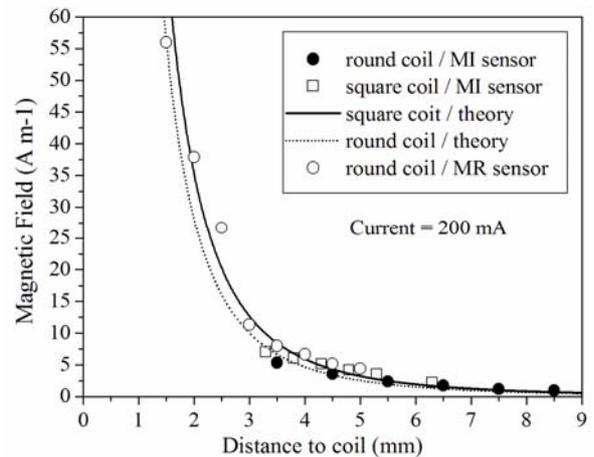

Figure 6: Measurement and simulation of the magnetic field generated by two 40 turns round and square coils as a function to the distance to coil center.

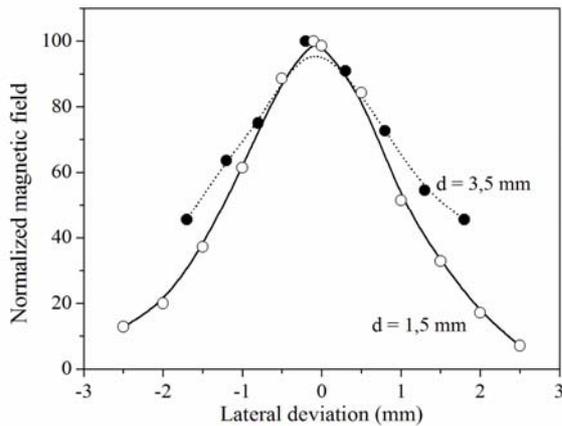

Figure 7: Measurement of the magnetic field generated by a 40 turns round coil as a function of the lateral deviation, at two different distances to the coil plane.

The measurements are presented on the figure 6. They have been compared to the theoretical values. The latter were computed with formulae (9) and (10) and averaged on 2 mm that is the length of the active part of the sensors. One can see that the measurements are in good agreement with this approximation.

The figure 7 presents the normalized values of the magnetic field in the planes located at 2 mm and 3 mm of the coil. According to these measurements, the magnetic field can be supposed as maximal (10 % deviation) above the whole surface of the coil.

## 3. APPLICATION TO MEMS ACTUATION

Figure 4 presents different considered strategies. The starting configuration (S1) supposes that the microcoils and the magnetically actuated devices are located on the opposite sides of the same silicon substrate. The distance between them is equal to the substrate thickness (280 μm for a 2 inches wafer). The configuration is optimal for injecting current through the track, due to the good thermal dissipation ability ('On-Si wafer' case, see fig 2).

This configuration is nevertheless not suitable if the technologies for processing the microcoils and the devices are not compatible. So it is considered the superposition of two substrates, the one supporting the microcoils and the other the actuated devices (S2). Compare to the configuration S1, the thermal dissipation ability remains important, but the distance between the microcoils and the devices is doubled.

If the flexibility of the support is required, microsystems can be dice-cutted from the configuration S1 or S2, depending on the process compatibility, then glued on a flexible film such as Kapton® (K1 and K2). Compare to S1 and S2, the distance between the microcoils and the actuated devices is not modified. However, the thermal dissipation ability is drastically reduced, as the microcoils are located between the silicon and the Kapton® film (worst than 'Microcoil on Kapton®' case).

To overcome this drawback, the microcoils can be directly processed on the Kapton® film (K3). This configuration is more favorable for current injection, but unfavorable because of the increasing of the distance between the microcoils and the actuated devices (compare to K2).

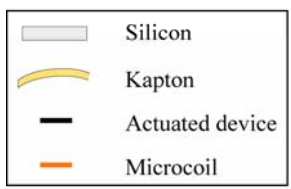
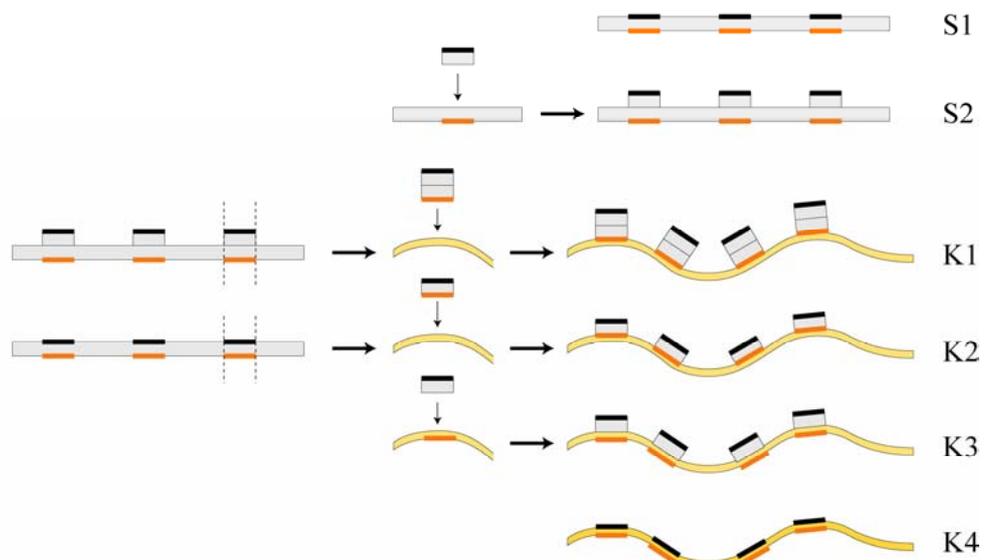

Figure 8: Different proposed configurations

*Johan MOULIN, Marion WOYTASIK, Emile MARTINCIC, Elisabeth DUFOUR-GERGAM*
*Copper Planar Microcoils Applied to Magnetic Actuation*

If the processes are compatible, these two components can be both processed on the Kapton® film (K4). This reduces considerably the distance between them and does not modify the maximal admissible current in the microcoil, compare to the configuration K3.

Using the maximal current that can be injected in the microcoils, it is possible to estimate the generated magnetic field corresponding to the previous configurations. The table 1 gives the estimation of the maximal current and the corresponding values of the magnetic field for a 40 turns round microcoil which track is 5 µm wide and 10 µw high. For calculation, the Kapton® film thickness is supposed to be 25 µm.

| Configuration | $I_{max}$ (mA) | Distance between coil and device (µm) | $H_{max}$ (A m$^{-1}$) |
|---|---|---|---|
| S1 | 300 | 280 | 6900 |
| S2 | 300 | 560 | 2055 |
| K1 | < 30 | 560 | < 205 |
| K2 | < 30 | 280 | < 690 |
| K3 | 30 | 305 | 610 |
| K4 | 30 | 25 | 2320 |

Table 1: Influence of the microcoil and actuated device configuration on the maximal current and the magnetic field. The microcoil is considered to be constituted by 40 turns of a 5 µm wide and 10 µm high track.

The results show that the silicon substrate is more efficient for magnetic field generation, compare to Kapton®. Indeed, the decreasing of the effective magnetic field due to the substrate thickness is largely compensated by the high value of the maximum current due to the thermal power dissipation ability of the silicon. Moreover, if flexibility is required, the optimal configuration corresponds to the configuration K4: the microcoils and the actuated devices are processed on the same film. In this case, the magnetic field reaches value close to 2400 A m$^{-1}$ (around 30 Oe).

## 4. CONCLUSION

The magnetic field generated at the center of microcoils with the same footprint has been calculated using analytical formulae. The results show that the more efficient configuration corresponds to the microcoil with the largest number of turns.

The maximum current allowed in the coil has been taken into account in order to define the Maximum Effective Magnetic Field at the center of the coil and the corresponding Joule losses. Their ratio represents the magneto-electrical efficiency of the device. Its value decreases slightly with the number of turns.

The measurement of the magnetic field generated by 40 turns round and square microcoils validated the simple used model.

Theses previous values have be implemented in different realistic configurations containing a microcoil and a magnetically actuated device. The largest magnetic influence if the largest if the microcoil and the magnetically actuated device are located on the opposite faces of a 280 µm-silicon wafer. However, large values can be reached if these components are separated by a 25 µm-flexible substrate as Kapton.